\documentstyle[epsf,aps,prl]{revtex}
\begin{document}
\title{Optical pumping
in dense atomic media: Limitations due to reabsorption 
of spontaneously emitted photons}
\author{ Michael Fleischhauer}
\address{Sektion Physik, 
Ludwig-Maximilians Universit\"at M\"unchen,
D-80333 M\"unchen, Germany}

\maketitle

\begin{abstract}
Resonant optical pumping in dense atomic media is discussed, 
where the absorption
 length is less than the smallest characteristic dimension of the sample.
It is shown that reabsorption and multiple scattering of spontaneous 
photons (radiation trapping) can substantially slow down the
rate of optical pumping. 
A very slow relaxation out of the target
state of the pump process is then sufficient to make optical pumping 
impossible. As model systems an inhomogeneously  and a radiatively
broadened 3-level system resonantly driven with a strong broad-band pump field
are considered. 
\end{abstract}

\pacs{42.50.Ct,32.80.Bx,32.80.Pj,42.68.Ay}

\section{Introduction}
Optical pumping is an established 
technique in atomic and molecular
physics to selectively populate or depopulate specific states or 
superpositions \cite{Kastler50,Cohen-Tannoudji75}. 
It is based on the absorption of photons of a 
specific mode and subsequent spontaneous emission into many modes. 
The dissipative nature of the latter part
makes it possible to transform  mixed into pure atomic states. 
From this results the importance of optical pumping
for state preparation in
systems with a thermal distribution of population and for 
laser cooling \cite{cooling}.

The maximum achievable rate of pumping is
determined by the escape time of the emitted photons, which in 
optically thin media is given by the free-space radiative
lifetime. When the medium becomes optically thick, however, i.e. when 
the absorption length becomes smaller than the smallest sample dimension,
the escape time of  photons can be substantially reduced.
This phenomenon, known as radiation trapping \cite{Holstein}, is due to 
reabsorption and multiple scattering of spontaneously emitted photons
and can drastically reduce the rate of optical pumping in 
dense media. These limitations could be of major importance in
many different fields as for instance near-resonance 
linear and nonlinear optics in dense media \cite{HI,NLO} or
the realisation of Bose condensation by velocity selective 
coherent population trapping (VSCPT) \cite{VSCPT-problem}.

To describe the reabsorption and multiple scattering of
photons we here utilize a recently developed approach  to
radiative interactions in dense atomic media \cite{Rad}.
In this approach a nonlinear and nonlocal single-atom density
matrix equation is derived which generalizes 
the linear theory of radiation trapping 
\cite{Holstein} to the nonperturbative regime. 
As a model system a 3-level $\Lambda$ configuration
driven by a strong broad-band field is considered and the limits of
(i) large inhomogeneous and (ii) purely radiative broadening are studied.

Let us consider  the $\Lambda$-type 
system shown in Fig.~1. A strong driving field with (complex) Rabi-frequency 
$\Omega(t)$ couples the lower state $|c\rangle$ to the excited state 
$|a\rangle$, which
spontaneously decays into $|c\rangle$ and 
$|b\rangle$. Since $|b\rangle$ is not coupled by the driving field,
this results 
in optical pumping from $|c\rangle$ to $|b\rangle$. 
We also take into account a possible finite lifetime of the 
target state  described 
by a population exchange between the lower states at rate 
$\gamma_0$.

\begin{figure}
\begin{center}
\leavevmode
\epsfxsize=6 true cm
\epsffile{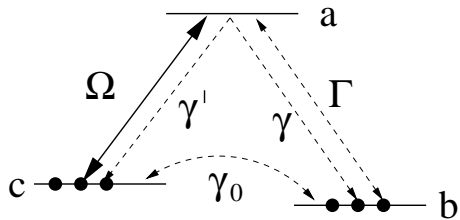}
\caption{Optical pumping in a $\Lambda$ system.}
\end{center}
\end{figure}

It was shown in  \cite{Rad} that the effect of the incoherent background
radiation can be described by additional (nonlinear and nonlocal)
pump and relaxation rates and level shifts in the 
single-atom density matrix equation.
If we assume orthogonal dipole moments or sufficiently different 
frequencies of the two optical transitions, the level shifts
are negligible. Also if the driving field is strong, the 
incoherent photons do not affect the pump transition 
$a \leftrightarrow c$. 
Thus we are left with a pump and decay rate $\Gamma(t)$ on the
$a \leftrightarrow b$ transition and the effective 
single-atom equations of motion
read in a rotating frame:
\begin{eqnarray}
{\dot\rho}_{aa} &=& -(\gamma+\gamma^\prime+\Gamma)\rho_{aa} +
\Gamma\rho_{bb} +i(\Omega^*\rho_{ac}-c.c),\label{rho_aa}\\
{\dot\rho}_{cc} &=& \gamma^\prime\rho_{aa} + 
\gamma_0\rho_{bb}-\gamma_0\rho_{cc}
-i(\Omega^* \rho_{ac}-c.c),\label{rho_cc}\\
{\dot\rho}_{ac} &=& -(i\Delta_{ac} +\Gamma_{ac})\rho_{ac} +i\Omega(\rho_{aa}
-\rho_{cc}).\label{rho_ac}
\end{eqnarray}
$\Delta_{ac}$ is the detuning of the drive field from resonance
and $\Gamma_{ac}$ is the respective coherence
decay rate. It should be noted, that $\Gamma$ is a function of 
the density matrix elements of all other atoms, and hence the
Eqs. (\ref{rho_aa}-\ref{rho_ac}) are 
nonlinear and nonlocal.

We are here interested in  genuine optical pumping  and therefore
consider a broad-band pump \cite{remark1}, i.e. $\Omega(t)$ is 
assumed to have a vanishing mean value and Gaussian $\delta$-like
correlations 
$\bigl\langle \Omega^*(t)\Omega(t^\prime)\bigr\rangle = 
R\, \delta(t-t^\prime).$
Formally intergating Eq.(\ref{rho_ac}), substituting the result back into 
Eqs.(\ref{rho_aa}) and (\ref{rho_cc}), and averaging over the Gaussian
distribution of the pump field leads to the rate equations
\begin{eqnarray}
{\dot\rho}_{aa} &=& -(\gamma+\gamma^\prime+\Gamma)\rho_{aa} +
\Gamma\rho_{bb} -R (\rho_{aa}-\rho_{cc}),\label{rate_aa}\\
{\dot\rho}_{cc} &=& \gamma^\prime\rho_{aa} + 
\gamma_0\rho_{bb}-\gamma_0\rho_{cc} +R (\rho_{aa}-\rho_{cc}).
\label{rate_cc}
\end{eqnarray}

\section{Collective decay rate}
We now have to determine the collective rate  
$\Gamma$.  $\Gamma$  is
proportional to the spectrum of the incoherent field at the position 
$\vec r_0$ and  the resonance frequency $\omega$ of
the atom under consideration \cite{Rad}
\begin{eqnarray}
\Gamma(\omega,t) =\frac{\wp^2}{\hbar^2}
{\widetilde D}(\vec r_0,\omega;t)
= \frac{\wp^2}{\hbar^2}\,
\int_{-\infty}^\infty\!\!\!d\tau\, \langle\langle {\hat E}^-(\vec r_0,t)
{\hat E}^+(\vec r_0,t+\tau)\rangle\rangle\, e^{i\omega\tau}.
\end{eqnarray}
Here ${\hat E}^\pm$ are the positive and negative frequency parts of the
field operators, $\wp$ is the dipole matrix element of the
atomic transition, and $\langle\langle AB\rangle\rangle \equiv
\langle AB\rangle-\langle A\rangle\langle B\rangle$. 
${\widetilde D}(\omega)$ can be obtained by summing
the spontaneous emission contributions of all atoms propagated through the
medium \cite{Rad}
\begin{equation}
D(1,1) = \int\!\!\!\int d3\, d4\, 
D^{\rm ret}(1,3)\,\Bigl(D^{\rm ret\,}(1,4)\Bigr)^*\, 
\Pi^{\, \rm s}(3,4).\label{GF_eq1}
\end{equation}
Here $D^{\rm ret}(1,2)$
is the retarded propagator of the electric field inside the medium, which
obeys a Dyson-equation in self-consistent Hartree approximation:
\begin{equation}
D^{\rm ret}(1,2) = D_{0}^{\rm ret}(1,2) -
\int\!\!\!\int d3\,d4\, 
 D_{0}^{\rm ret}(1,3)\, \Pi^{\rm ret}(3,4)\, 
D^{\rm ret}(4,2).
\label{GF_eq2}
\end{equation}
In Eqs.(\ref{GF_eq1}) and (\ref{GF_eq2}) the numbers $1,2\dots$ stand for
$\{\vec r_1,t_1\},\{\vec r_2,t_2\}\dots$, and the intergrations 
extend  over time from $-\infty$
to $+\infty$ and over the whole sample volume. 
$D_0^{\rm ret}$ is the free-space retarded propagator of the
electric field. 
For simplicity we here have disregarded polarisation.
We also have introduced the atomic source correlation
\begin{equation}
\Pi^{\, \rm s}(\vec r_1,t_1;\vec r_2,t_2) =
\frac{\wp^2}{\hbar^2}\sum_j
\bigl\langle\bigl\langle \sigma_j^\dagger(t_1) \sigma_j
(t_2)\bigr\rangle\bigr\rangle\, \delta(\vec r_1-\vec r_j)\,
\delta(\vec r_2-\vec r_j)\label{source}
\end{equation}
and the atomic response function 
\begin{equation}
\Pi^{\rm ret}(\vec r_1,t_1;\vec r_2,t_2) =
\frac{\wp^2}{\hbar^2}\,
\Theta(t_1-t_2)\sum_j 
\bigl\langle \bigl[\sigma_{j}^\dagger(t_1), \sigma_{j}
(t_2)\bigr]\bigr\rangle\, \delta(\vec r_1-\vec r_j)\,
\delta(\vec r_2-\vec r_j),\label{response}
\end{equation}
where $\sigma_j=|b\rangle_{jj}\langle a|$ is the spin-flip operator
of the $j$th atom and $\Theta$ is the Heaviside step function. 
In terms of the $\sigma$'s the dipole operator of the $j$th atom
reads $d_j=\wp(\sigma_j + \sigma_j^\dagger)$.
The names reflect the physical meaning of the quantities (\ref{source},
\ref{response}).
The Fourier-transform of $\Pi^{\, \rm s}$ is proportional to
the spontaneous emission spectrum of the atoms and that of $\Pi^{\rm ret}$
gives the susceptibility of the medium. 
Eqs.(\ref{GF_eq1}) and (\ref{GF_eq2}) represent a nonperturbative summation
of the spontaneous radiation contributions of all atoms propagated through the
medium. It assumes a Gaussian statistics, which is however a good 
approximation for the background radiation.

The Dyson-equation (\ref{GF_eq2}) was solved in \cite{Rad} 
with some approximations in a macroscopic (continuum) limit
where  $\Pi(\vec r_1,t_1;\vec r_2,t_2) = 
\int d^3\vec r\, 
P(\vec r,t_1,t_2)\, \delta(\vec r_1-\vec r)\, \delta(\vec r_2 -\vec r)$.
This yielded for the collective decay rate
\begin{equation}
\Gamma(\omega;t) = 
\frac{\wp^2 \omega^4 }{(6 \pi)^2 \epsilon_0^2 c^4}
\int_V\! d^3\vec r \, \frac{e^{2 q_0^{\prime\prime}(\vec r,\omega;t) r}}{r^2}
\, {\widetilde P}^{\rm \, s} (\vec r,\omega;t),
\label{G_sol}
\end{equation}
where $r=|\vec r_0-\vec r|$ is the distance bewteeen the source and the
probe atom. The probability that a photon reaches the
probe atom is determined by the absorption coefficient 
\begin{equation}
q_0^{\prime\prime}(\vec r,\omega,t) =
\frac{\hbar \omega}{3\epsilon_0 c} \, {\rm Re}\, \left[{\widetilde P}^{\rm ret}
(\vec r,\omega;t)\right].
\end{equation}

One can easily calculate the atomic  source and response functions 
for the $\Lambda$-system of Fig.~1.  
\begin{eqnarray}
{\widetilde P}^{\rm ret}(\vec r_j,\omega,t)&=& \frac{\wp^2}{\hbar^2}
N\, \overline{\, \frac{\rho_{aa}^j(t) -\rho_{bb}^j(t)}{\Gamma_{ab}
+i(\omega-\omega_{ab}^j)}\, },\label{Pi_ret_TLA}\\
&&\nonumber\\
{\widetilde P}^{\rm\, s}(\vec r_j,\omega,t)&=& \frac{2\wp^2}{\hbar^2}
N\overline{\, \frac{\rho_{aa}^j(t)\Gamma_{ab}}{(\Gamma_{ab})^2
+(\omega-\omega_{ab}^j)^2}\, },\label{Pi_s_TLA}
\end{eqnarray}
where $N$ is the density of atoms, $\omega_{ab}^j$ is the resonance frequency
of the $j$th atom, $\Gamma_{ab}$ the coherence decay rate of the 
corresponding transition 
and the overbar denotes averaging over a
possible inhomogeneous distribution of frequencies.

At this points we shall distinguish two limiting cases. We first consider
the limit of large Doppler-broadening 
and secondly the case of purely radiative broadening. 

\section{Inhomogeneously broadened system}
The approach of \cite{Rad} is 
based on the Markov approximation of a spectrally  broad 
incoherent radiation. 
This approximation is justified for example in an inhomogeneously
broadened system. We therefore discuss first the case of large 
Doppler-broadening. If we are interested
in the population 
dynamics on a time scale slow compared to velocity changing
collissions, we may set $\rho_{\mu\mu}^j(t)=\overline{\, \rho_{\mu\mu}^j(t)\, }
\equiv \rho_{\mu\mu}(\vec r_j,t)$ and thus have the same population dynamics
in all velocity classes. Since $\Gamma$ depends on the
populations of all atoms, Eqs.(\ref{rate_aa}) and (\ref{rate_cc})
are nonlocal. 
 In the case of a constant density of atoms and
a homogeneous pump field, $\Gamma$ and hence all density matrix elements
will be approximately homogeneous. 
We therefore make a simplifying approximation and 
disregard the space dependence. The volume integral
is then carried out by placing the probe atom in the center
of the sample.
This yields for a Gaussian Doppler-distribution of 
width $\Delta_D\gg \gamma$
\begin{equation}
\frac{\Gamma(\omega,t)}{\gamma}=
 \frac{\rho_{aa}(t)}{\rho_{bb}(t)-\rho_{aa}(t)}
\left[ 1-\exp\left(-H(t) e^{-\Delta^2/2\Delta_D^2}\right)\right],
\label{Gamma_omega}
\end{equation}
where $\Delta=\omega-\omega_{ab}^0$ is the detuning from
the atomic resonance at rest, and 
$ H(t)=K \, [\rho_{bb}(t)-\rho_{aa}(t)].$
$K=g\, N \lambda^2 d_{\rm eff}$ with $g=\gamma/\sqrt{2 \pi}\, \Delta_D$
characterizes the number of atoms within one relevant velocity class
in a volume given by the wavelength squared
and the effective escape distance $d_{\rm eff}$. In deriving
(\ref{Gamma_omega})  we
have used the relation between the free-space radiative 
decay rate $\gamma$ and the dipole moment $\wp$: 
$\wp^2 = 3\pi\hbar\epsilon_0 c^3 \gamma/\omega^3$
\cite{Louisell}.
$d_{\rm eff}$ corresponds for a long cylindrical slab  to the
cylinder radius; for a thin disk to its thickness
and for a sphere to its radius.

Averaging over the inhomogeneous velocity distribution of the atoms
eventually yields
\begin{eqnarray}
\Gamma(t)=\overline{\, \Gamma(\omega,t)\, }
 &=&
 \int_{-\infty}^\infty \!\!\!d\omega\, \frac{1}{\sqrt{2\pi}\Delta_D}
e^{-\Delta^2/2\Delta_D^2} \, \Gamma(\omega,t)\nonumber\\
&=& \gamma\, \frac{\rho_{aa}(t)}{\rho_{bb}(t)-\rho_{aa}(t)}
\frac{1}{\sqrt{\pi}}\, \int_{-\infty}^\infty\!\!\! dy\, e^{-y^2}
\left[ 1-\exp\left(-H(t) e^{-y^2}\right)\right].\label{G_TLA_approx}
\end{eqnarray}

In Fig.~2a we have shown the population in the target state $|b\rangle$
as function of time starting from equal populations of 
levels $|c\rangle$ and $|b\rangle$ at $t=0$.
We here have assumed that the target
state is stable, i.e. $\gamma_0=0$. One recognizes that optical
pumping is considerably slowed down already for
values of $K$ on the order of 10, which usually corresponds to much less
than one atom per $\lambda^3$. The slow-down of pumping
 is further illustrated  in Fig.~2b,
where the effective pump rate defined as
\begin{equation}
\Gamma_{\rm p}\equiv 
- {\frac{d}{dt}}{\rm ln}[\rho_{aa}+\rho_{cc}]
\end{equation}
is plotted normalized to the value in an optically
thin medium ($\Gamma_{\rm p}^0=\gamma/2$). 
One can see that the optical pump rate approaches a constant
asymptotic value, which for $K\gg 1$ and large pump rates $R$ is given by
\begin{equation}
\Gamma_{\rm p}^{\rm as}= 
\frac{\gamma}{2 K \bigl(\pi\, {\rm ln} K\bigr)^{1/2}}\ll\frac {\gamma}{2}.
\end{equation}

\begin{figure}
\begin{center}
\leavevmode
\epsfxsize=14.3 true cm
\epsffile{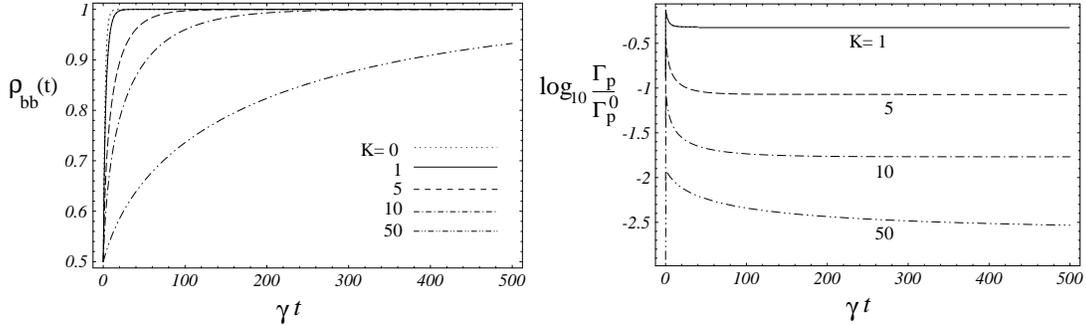}
\caption{(a) Time evolution of 
population of level $|b\rangle$ for $\gamma_0=0$,
$R/\gamma=10$ and $\gamma^\prime/\gamma=1$ for different density
parameters $K=g\, N\lambda^2 d_{\rm eff}$,
$g=\gamma/\sqrt{2\pi}\Delta_D$. 
(b) Corresponding effective rate of optical pumping}
\end{center}
\end{figure}

Since we have assumed in the plots of Fig.~2 
an infinitely long-lived target state ($\gamma_0=0$),
all populations eventually ends up in $|b\rangle$.
However if $\gamma_0$ is nonzero and in particular if it becomes comparable to
the asymptotic rate $\Gamma_p^{\rm as}$, the steady-state populations of all
states equalize. In this case optical pumping is less and less efficient
and becomes eventually impossible. 
This is illustrated in Fig.~3, where the stationary population
in state $|b\rangle$ is shown as a function of the density parameter $K$
for different values of $\gamma_0$.

\begin{figure}
\begin{center}
\leavevmode
\epsfxsize=7 true cm
\epsffile{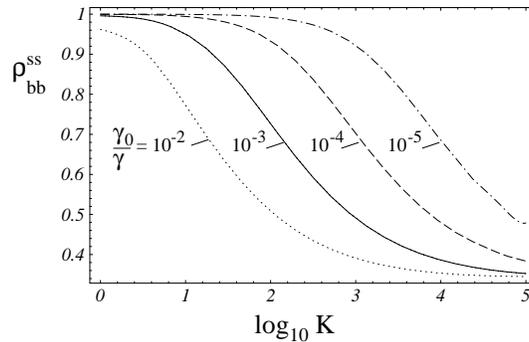}
\caption{Stationary population in level $|b\rangle$ for
$R/\gamma=10$, $\gamma^\prime/\gamma=1$ and different values of $\gamma_0$
as function of density parameter $K$.}
\end{center}
\end{figure}

\section{Radiatively broadened system}
We now discuss the case of a radiatively broadened system. 
In analogy to the case of inhomogeneous
broadening, we find for the spectral distribution
\begin{equation}
\frac{\Gamma(\omega,t)}{\gamma}=
 \frac{\rho_{aa}(t)}{\rho_{bb}(t)-\rho_{aa}(t)}
\left[1-\exp\left(-H(t)
\frac{\gamma_{ab}\Gamma_{ab}}
{\Gamma_{ab}^2+\Delta^2}\right)\right],\label{G_rad}
\end{equation}
where $\Delta=\omega-\omega_{ab}$, $\Gamma_{ab}=\gamma_{ab}+\Gamma$,
and $\gamma_{ab}=(\gamma+\gamma^\prime+R+\gamma_0)/2$, and
$ H(t)={\widetilde K} \, [\rho_{bb}(t)-\rho_{aa}(t)].$
Here ${\widetilde K}={\widetilde g}\, N\lambda^2 d_{\rm eff}$ with 
${\widetilde g}=\gamma/2\pi \gamma_{ab}$.
As opposed to the corresponding relation in the inhomogeneous case,
Eq.(\ref{G_rad}) determines the collective decay rate only implicitly,
and $\Gamma$ needs to be calculated self-consistently. 
For small atomic densities or $\rho_{aa}\approx\rho_{bb}$ the exponential
function in Eq.(\ref{G_rad}) can be expanded into a power series.
The first nonvanishing term found from this has the same
spectral shape than the single-atom response function. 
In such a case the Markov approximation used in \cite{Rad} is
no longer valid and the approach is quantitatively incorrect. 
We shall nevertheless use it and discuss the
range of validity afterwards. 

We find that in the case of radiative
broadening the rate of optical pumping decreases exponentially with the
density parameter as opposed to $[N\lambda^2 d_{\rm eff}]^{-1}$ 
in the inhomogeneous case.
For sufficiently large pump rates $R$ and stable target state ($\gamma_0=0$)
the asymptotic rate of optical
pumping is here
\begin{equation}
\Gamma_{\rm p}^{\rm as}=\frac{\gamma}{2}\exp\bigl\{-{\widetilde K}\bigr\}.
\end{equation}
Physically this is due to the fact that here the incoherent photons are
in resonance with all atoms, which drastically increases the
scattering probability. As a consequence much smaller decay rates
$\gamma_0$ out of the target state are sufficient to make optical
pumping impossible. This is illustrated in Fig.~4, where we have plotted 
the stationary population in state $|b\rangle$ as function of the
density parameter $K_0=N \lambda^2 d_{\rm eff} /2\pi$ 
for different values of $\gamma_0$.

\begin{figure}
\begin{center}
\leavevmode
\epsfxsize=7 true cm
\epsffile{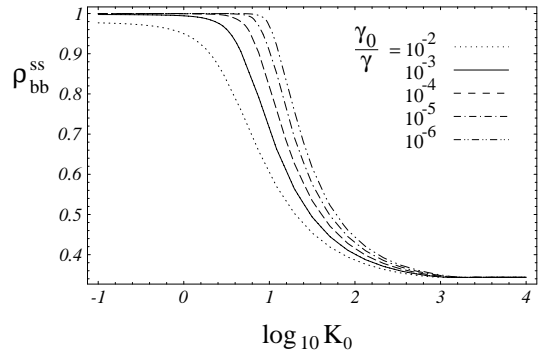}
\caption{Same as Fig.3 for radiatively broadened system;
$ K_0=
N \lambda^2 d_{\rm eff}/2\pi$, 
$R/\gamma=10$, $\gamma^\prime/\gamma=1$}
\end{center}
\end{figure}

In order to check the validity of the Markov approximation, we have
shown in Fig.~6 the stationary normalized spectral distribution 
$\Gamma(\omega)/\Gamma$ for $K_0=1$, $10$ and $100$ and $\gamma_0/
\gamma=10^{-4}$. Also plotted is the atomic absorption spectrum for $K_0=1$
(solid line).

\begin{figure}
\begin{center}
\leavevmode
\epsfxsize=6.5 true cm
\epsffile{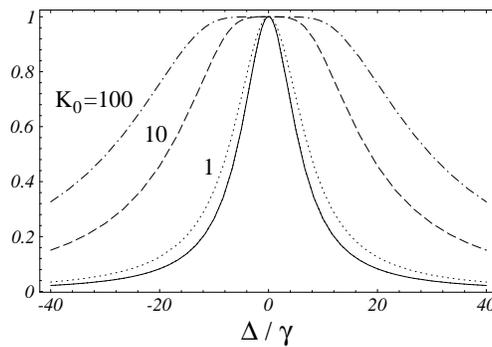}
\caption{Spectral distribution of incoherent background radiation
for $R/\gamma=10$, $\gamma^\prime=\gamma$, $\gamma_0/\gamma=10^{-4}$,
and $K_0=1$ (dotted), $K_0=10$ (dashed) and $K_0=100$ (dashed-dotted).
Also shown is the normalized absorption spectrum for $K_0=1$. 
}
\end{center}
\end{figure}

\noindent One recognizes that spectrum of the background radiation has only a 
slightly larger width than the atomic response for $K_0=1$. In this
case the Markov approximation is not valid.
The situation  however improves 
when the density is increased. Thus Fig.~4 has only qualitative character
for lower densities.

\ 

\section{Summary}
We have shown that resonant optical pumping in a dense atomic medium 
is substantially different from optical puming in dilute systems.
When the absorption length of spontaneously emitted photons 
process becomes less than the minimum escape distance,
these photons are trapped inside the medium and cause repumping
of population.
This leads to a considerable slow-down of the transfer rate and 
can make optical pumping  impossible if the target state
of the pump process has a finite lifetime. The effect
is much less pronounced in inhomogeneously broadened systems due to
the reduction of the spectral density of background photons.

These results may have some important consequences.  
It is practically impossile to use resonant optical pumping in media with 
$N\lambda^3\sim 1$. This sets strong limits to the possibility to prepare
pure states or coherent superpositions in systems with 
initial thermal occupation of states, such as Hyperfine ground levels 
of alkali at room temperature. Even though the above analysis did not
take into account quantum properties of the atoms and considers
only resonant pumping, the results indicate, that
it may be very difficult to achieve Bose Condensation via VSCPT in optical
lattices \cite{lattices}. 
Also the present results show that electromagnetically induced 
transparency (EIT) \cite{EIT} in dense media cannot be understood 
as the result of optical puming into a dark state.
Essential for EIT in dense media is an entirely coherent evolution
\cite{Lu} via stimulated adiabatic Raman passage \cite{STIRAP}.
Some of these aspects will be discussed in more detail elsewhere.

\section*{Acknowledgement}

The author would like to thank C.M. Bowden and
S.E. Harris for stimulating discussions.


\begin{references}
\bibitem{remark1} A narrow-bandwidth coherent laser would
cause Autler-Townes splitting and other interference effects
in the atomic response and spontaneous emission spectrum. Furthermore
in dense media the Lorentz-Lorenz corrections to the mean amplitude would
have to be taken into account as well.
\end{references}
\end{document}